\documentclass[aps,prb,showpacs,twocolumn,reprint,superscriptaddress]{revtex4-2}
\usepackage{amsmath}
\usepackage{amssymb}
\usepackage{graphicx}
\usepackage{color}
\usepackage{multirow}
\usepackage{dcolumn}
\usepackage{textcomp}
\usepackage{longtable}
\usepackage{makecell}
\begin{document}

\title{Combining intrinsic and sliding-induced polarizations for multistates in two dimensional ferroelectrics}

\author{Chuhan Tang}
\affiliation{Key Laboratory for Matter Microstructure and Function of Hunan Province,
Key Laboratory of Low-Dimensional Quantum Structures and Quantum Control of Ministry of Education, School of Physics and Electronics, Hunan Normal University, Changsha 410081, China}

\author{Zhiqiang Tian}
\affiliation{Key Laboratory for Matter Microstructure and Function of Hunan Province,
Key Laboratory of Low-Dimensional Quantum Structures and Quantum Control of Ministry of Education, School of Physics and Electronics, Hunan Normal University, Changsha 410081, China}

\author{Tao Ouyang}
\affiliation{Hunan Key Laboratory for Micro-Nano Energy Materials and Device and School of Physics and Optoelectronics, Xiangtan University, Xiangtan 411105, Hunan, China}

\author{Anlian Pan}
\affiliation{Key Laboratory for Matter Microstructure and Function of Hunan Province,
Key Laboratory of Low-Dimensional Quantum Structures and Quantum Control of Ministry of Education, School of Physics and Electronics, Hunan Normal University, Changsha 410081, China}
\affiliation{Key Laboratory for Micro-Nano Physics and Technology of Hunan Province, 
College of Materials Science and Engineering, Hunan University, Changsha 410082, China}

\author{Mingxing Chen}
\affiliation{Key Laboratory for Matter Microstructure and Function of Hunan Province,
Key Laboratory of Low-Dimensional Quantum Structures and Quantum Control of Ministry of Education, School of Physics and Electronics, Hunan Normal University, Changsha 410081, China}
\affiliation{State Key Laboratory of Powder Metallurgy, Central South University,  Changsha 410083, China}

\email{mxchen@hunnu.edu.cn}

\date{\today}

\begin{abstract}
Going beyond the bistability paradigm of the charge polarizations in ferroelectrics is highly desired for ferroelectric (FE) memory devices toward ultra-high-density information storage.  Here, we propose to build multistates by combining the intrinsic and sliding-induced polarizations. The physics is that there is at least one order of magnitude difference in the energy barriers between these two types of polarization, which leads to a significant difference in the electric fields for reversing the polarization. This difference, along with the symmetry breaking, allows for a unique flipping mechanism involving layer-by-layer sliding followed by layer-by-layer flipping during the transformation of the multistates. As a result, six and ten switchable states can be achieved for the 1T" bilayers and trilayers, respectively. We further illustrate the concept in H-stacking bilayers and trilayers of 1T" transition-metal dichalcogenides by first-principles calculations.  Our study provides a new route to design novel polarization states for developing next-generation memory devices. 
\end{abstract}

\keywords{Transition-metal dichalcogenide bilayer; ferroelectrics; multistates}

\maketitle

\section{Introduction}
Ferroelectrics have promising applications in nonvolatile memory devices with high storage density, fast read/write speed, and low power consumption~\cite{JScott2007, LMartin2016}. There has been a growing demand for continuously increasing the storage density of FE devices in the era of big data. For this purpose, the thickness of the ferroelectrics has to be reduced. Unfortunately, conventional perovskite ferroelectrics suffer from the problem of the ferroelectricity being suppressed as the film thickness decreases to a critical value due to the depolarization field~\cite{junquera2003, Fong2004, dawber2005}.

Two-dimensional (2D) ferroelectrics have advantages in addressing the scaling issue since they have no dangling bonds at the surfaces. To date, a great number of 2D ferroelectrics have been experimentally identified and theoretically
predicted~\cite{chang2016,fei2016,higa2020,chang2020,guan2022,Ma2021,Ding2017,Zhou2017,Cui2018,Gou2023}. 
For instance, it was experimentally shown that the ultrathin films of rhombohedral SnTe exhibit unexpectedly much higher Curie temperatures than the corresponding bulk phase ~\cite{chang2016}.  Theoretical calculations predicted that monolayers of the orthorhombic group-IV monochalcogenides also exhibit ferroelectricity with in-plane polarizations and high Curie temperatures~\cite{fei2016}.  
Interestingly, a few groups of 2D ferroelectrics with out-of-plane polarizations (OOP), e.g.,  In$_2$Se$_3$ and CuMP$_2$S$_6$ (M = In and Cr), were recently found by both theoretical predictions and experiments~\cite{Ding2017,Zhou2017,Cui2018,Belianinov2015,Liu2016,Io2023,Ma2023, Wang2023}.   
Moreover, ferroelectricity with the OOP was also found in the distorted 1T phases of transition-metal chalcogenides (TMDs)~\cite{Shirodkar2014,yuan2019,Choi2020,lipatov2022}. Such ferroelectricity is highly desired for their applications in FE devices.

In addition to the intrinsic 2D ferroelectricity, it was recently shown that stacking nonpolar monolayers into bilayers and multilayers can also give rise to simultaneous polarization~\cite{Lli2017,Zfei2018,WuM2021,vizner2021,Xwang2022,Ywan2022,Ji2023,YangPRL2023}. This property is due to the electronic reconstruction caused by stacking-induced symmetry breaking, for which the interlayer sliding is responsible for the reversal of the polarization.  

In this study, we propose to design polarization multistates by combining the intrinsic and sliding ferroelectricity in two-dimensional ferroelectrics.  These systems have one more degree of freedom for switching the polarizations than any of the individual counterparts. We further illustrate the concept in bilayers and trilayers of  1T"-TMD by performing first-principles calculations. We find that six and ten states can be achieved for the bilayer and trilayer respectively by properly controlling the external electric field. Our study provides a new way to design novel polarization states for ultrahigh-density storage technology beyond conventional FE devices.

\section{Results and discussion}

\subsection{Model and concept}
\begin{figure}
 \includegraphics[width=.95\linewidth]{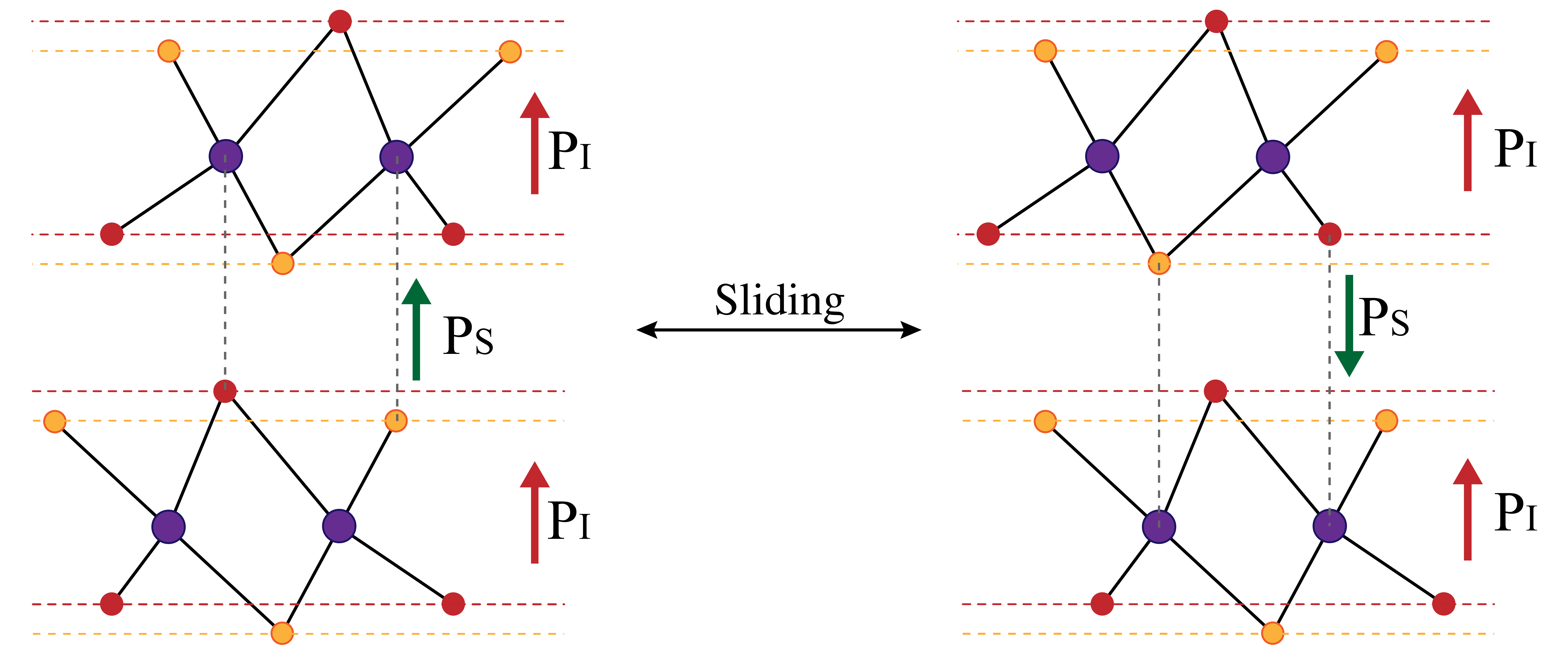}
  \caption{Sketch of polarization multistates in composite ferroelectrics, which exhibit both intrinsic and stacking-induced polarizations. The former ($P_I$) originates from the motion of ions within the layers, while the latter ($P_S$) is dependent on specific stacking. There are eight polarization states for their combination in a bilayer. (a, b) show two configurations of interlayer polarization coupling for a bilayer with distorted 1T phase (ferroelectric) in the H-stacking.}
 \label{fig1}
\end{figure}

Our concept makes use of the intrinsic polarizations ($P_I$) due to ionic motions and those induced by stacking ($P_S$), which is depicted in Fig.~\ref{fig1}.  For intrinsic 2D ferroelectrics such as CuInP$_2$S$_6$ and In$_2$Se$_3$, their bilayers have four different polarization configurations~\cite{DingPRL2021,tawfik2018van}. Among them, two have the interlayer FE coupling with opposite net polarization. The other two have interlayer antiferroelectric (AFE) couplings, which are in the tail-to-tail and head-to-head configurations, respectively.  For the sliding ferroelectricity, the polarization is dependent on the stacking order. For instance, for the $H$ phase of TMD layers without inversion, R-stacking gives rise to net electronic polarizations, which is absent for the H-stacking. However, the $T$ phase requires the H-stacking for nonvanishing polarization.  In both stackings that induce polarizations, the metal atoms in one layer sit above the nonmetal atoms in the other TMD layer. This property gives rise to the bistability nature of the stacking-induced polarizations $P_S$ for the bilayers. However, combining these two in one system can give rise to more states for their thin films. Fig.~\ref{fig1} shows that there are octuple polarization states for a bilayer with both the intrinsic and stacking-induced polarizations.   

One important feature of this type of ferroelectrics is that the difference in energy barriers for the two types of polarization allows for a unique switching that combines layer-selective flipping and sliding by controlling external electric fields. The energy barrier for the intrinsic spontaneous polarizations due to ionic motions is generally much larger than that of the interlayer sliding.  For instance, it is about 66 meV/f.u. for In$_2$Se$_3$ monolayer~\cite{Ding2017} and about 218 meV/f.u.  for CuInP$_2$S$_6$ monolayer~\cite{GLYu2021}. Whereas the energy barrier for layer sliding in the bilayer of h-BN is only about 9 meV/f.u.~\cite{Lli2017}.  Thus, there is at least one order of magnitude difference in the energy barriers between the two different polarizations. As a result, the external electric fields for switching them can be dramatically different. Namely, small electric fields are needed to switch $P_S$ and much larger electric fields are required to flip $P_I$.  Thus, the two types of polarizations in such composite ferroelectrics can be manipulated by properly controlling the electric field.

\subsection{Material realization}
\begin{figure*}
\centering
\includegraphics[width=.95\linewidth]{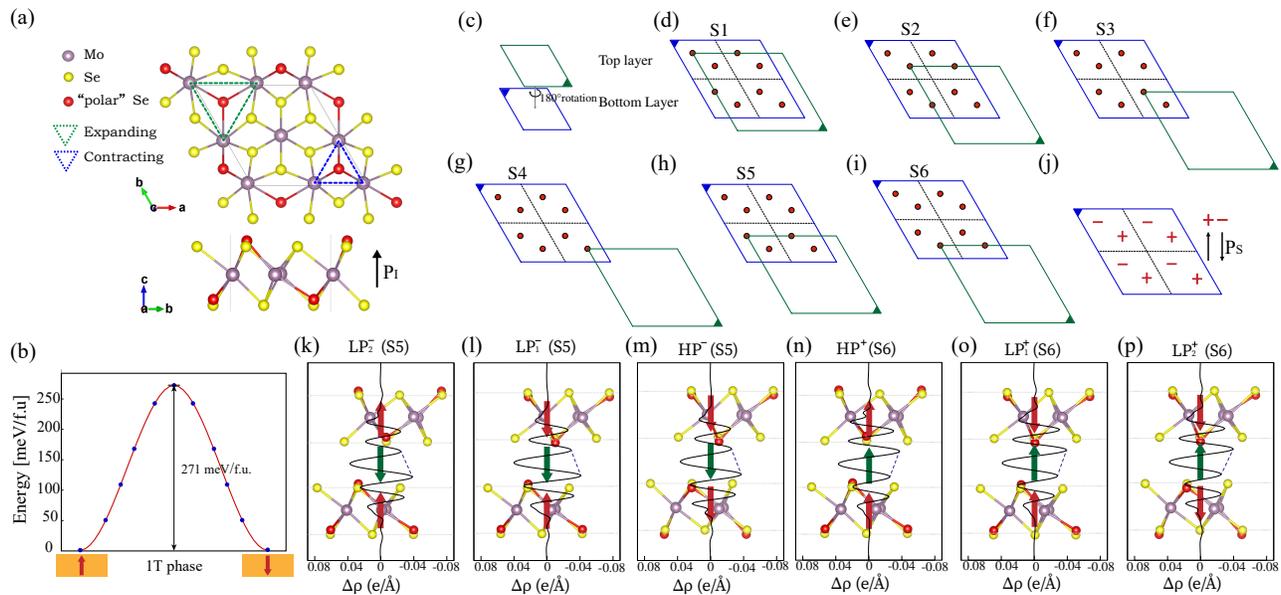}
\caption{Polarization multistates in a 1T"-MoSe$_2$ bilayer.  (a) Geometric structure of the 1T"-MoSe$_2$ monolayer, which is derived from a 2 $\times$ 2 supercell of the 1T phase. (b) The energy barrier in meV per formula unit(f.u.) for reversing the polarizations in the 1T"-MoSe$_2$ monolayer. (c-i) The stacking configurations of the H-stacking 1T"-TMD bilayer.  The cells for the 1T phase are denoted by dashed parallelograms.  (j) The signs of $P_S$ for the stackings show in (d-i). (k-l) Polarizations for S5 and S6 in different interlayer couplings.  $P_I$ and $P_S$ are denoted by red and green arrows, respectively.  $\Delta \rho (\mathrm{e/Å})$ denotes the planar-averaged charge density difference. LP and HP denote low- and high polarization states, respectively.}
\label{fig2}
\end{figure*}

We find that the bilayers of distorted TMD in the d1T and 1T" phases by appropriate
  stackings are potential candidates for the proposed multistates. These phases are derived from a $\sqrt{3} \times \sqrt{3}$ and a 2 $\times$ 2 structural reconstructions of the 1T phase, respectively. In these phases, the transition-metal atoms have in-plane displacements.  For instance, for the 1T" phase three of them form a contracted triangle, which also leads to an expanded triangle (see Fig.~\ref{fig2}a). As a result, the chalcogen atoms sitting on the hollow site of the contracted triangles have an out-of-plane displacement relative to those on the hollow sites of the expanded ones. Such displacements give rise to the out-of-plane polarizations for the d1T and  1T"-TMD phases, which were already confirmed by experiments~\cite{yuan2019,lipatov2022}. The energy barriers for the FE transforming from the nudged elastic band (NEB) calculations~\cite{henkelman2000,henkelman2000improved} are 212, 271, and 306 meV/f.u. for 1T"-MoS$_2$, MoSe$_2$, and MoTe$_2$, respectively. 

Despite the structural reconstruction, the stacking order responsible for sliding ferroelectricity in the bilayer of the $T$ phase still appears in the corresponding stacking of the 1T" bilayers. Therefore, stacking-induced polarizations appear in the 1T"-TMD bilayers with H-stacking. Like the standard sliding ferroelectrics, proper shiftings of the 1T"-TMD layers lead to the reverse of the stacking-induced polarization. 

To validate our idea, we have performed first-principles calculations for the H-stacking bilayers of 1T"-MoSe$_2$ using the Vienna Ab initio Simulation Package~\cite{kresse1996}. The exchange-correlation functional is parametrized using the Perdew–Burke–Ernzerhof (PBE) formalism in the generalized gradient approximation~\cite{PBE1996}.
The pseudopotentials were constructed by the projector augmented wave method\cite{bloechl1994,kresse1999}.
Van der Waals dispersion forces between the adsorbate and the substrate 
were accounted for by using the DFT-D3 method~\cite{grimme2010}.
A 12 $\times$ 12 Monkhorst-Pack $k$-mesh was used to sample the two-dimensional Brillouin zone and a plane-wave energy cutoff of 400 eV was used for structural relaxation and self-consistent calculations. 

The polarization was calculated using the Berry phase method~\cite{Resta1992,Vanderbilt1993}, which gives 0.32 pC/m, 0.18 pC/m, and 0.22 pC/m, for 1T"-MoS$_2$, MoSe$_2$, and MoTe$_2$. The result for MoS$_2$ agrees with a previous study~\cite{lipatov2022}. In addition, to compare with the polarization of three-dimensional (3D) ferroelectrics, we also calculated the polarization by defining the thickness of the monolayers based on the range of planar averaged potential along the out-of-plane direction of the slab~\cite{Boris2025}. The calculated polarization is 0.018 $\mu$C/cm$^2$ for 1T"-MoSe$_2$, which is much smaller than that for the bulk phase of BaTiO$_3$~\cite{Ghosez2022}, but is comparable to that for the bilayer of InSe~\cite{WuM2021}. Moreover, we have also performed the \textit{ab initio} molecular dynamics simulations for the monolayer and bilayer of 1T"-MoSe$_2$. The results show that their geometric structures preserve well at room temperature and the polarization is also maintained [see Sec. S1 in Supplemental Material (SM)]. 

Systematic calculations were performed for the H-stacking 1T"-MoSe$_2$ bilayer. Figs.~\ref{fig2}(a) and (b) show the geometric structure for a 1T"-MoSe$_2$ monolayer and the energy barrier for switching the polarization, respectively. Because the unit cell of the 1T"-TMD is four times that of the 1T phase, there are eight stacking configurations [Figs.~\ref{fig2}(c-j)] for its bilayer. However, only six of them are nonequivalent considering the geometric symmetry. We have performed calculations for the system with various interlayer and intralayer polarization states. Our calculations find that they are all energy local minimums. Among them, S2 has the lowest energy for the state with both interlayer and intralayer ferroelectric polarization coupling, i.e., $\frac{\uparrow}{\uparrow}$, where the intralayer polarizations are shown by arrows. Experimentally, this state in principle can be achieved by polling. The energy differences between various stacking configurations are small ($\sim$ 20  meV/f.u.). The stacking-induced polarizations $P_S$ are calculated by subtracting those of the individual layers from the total polarization. Our results find that the sign of $P_S$ for each stacking remains unchanged regardless of the interlayer polarization coupling (IPC). The signs of $P_S$ for the eight configurations are shown in Fig.~\ref{fig2}j, which can be classified into two groups with opposite signs, i.e., $\{$S1, S3, S5$\}$ and $\{$S2, S4, S6$\}$. 

Below we only show three of them for S5 and S6 (see Fig.~\ref{fig2}). The one with tail-to-tail interlayer coupling is not shown since it is absent for the polarization switching as discussed below. The calculated polarizations are listed in Table ~\ref{table1}.  There are only slight changes in the magnitude of $P_S$ with respect to different IPCs for each stacking. This property can be understood since the stacking order remains unchanged, although the IPC varies. Therefore, the electronic reconstruction is preserved. We have performed an analysis of the differential charge density ($\Delta \rho$) by subtracting the charge density of the individual layers from that of the bilayer [Figs.~\ref{fig2}(k-p)]. The results confirm the electronic origin and clearly explain the signs of $P_S$ for different stacking. Note that the intrinsic polarization of the 1T"-TMD monolayer is small and both the intrinsic and stacking-induced polarizations depend on the structural and electronic properties of materials. Therefore, the stacking-induced polarization can be stronger than the intrinsic polarization for the 1T"-TMD bilayers.  Moreover, the stacking also gives rise to in-plane polarization. However, it is strongly dependent on the stacking configuration and interlayer polarization coupling (see Sec. S2 in SM), which can be manipulated by the layer sliding using the out-of-plane electric fields. Therefore, we focus on the transformation of the OOP polarization below.
\begin{table}[!ht]
\centering
\caption{Polarization states for S5 and S6.  $P_I^T$ and  $P_I^B$ ($pC/m$)  denote the intrinsic polarizations in the top and bottom layers, respectively. $P_I^T$-$P_I^B$ represent the interlayer polarization configurations. $P_{tot}$ denotes the total polarization ($pC/m$).}
\label{table1}
\begin{tabular}{p{0.8cm}<{\centering}p{2cm}<{\centering}p{1cm}<{\centering}p{1cm}<{\centering}p{1cm}<{\centering}p{1cm}<{\centering}}
    \hline
    \hline
        Stacking          & $P_I^T$-$P_I^B$                   & $P_{tot}$ & $P_I^T$ & $P_I^B$ & $P_S$  \\ \hline
 \multirow {4}*{S5}  & $\uparrow \uparrow$           & -0.18      & 0.18           & 0.20            & -0.56 \\ 
                                 & $\downarrow \downarrow$  & -0.69     & -0.23         & -0.17           & -0.29 \\ 
                                 & $\uparrow \downarrow$       & -0.48     & 0.16            & -0.17           & -0.46  \\ 
                                 & $\downarrow \uparrow$       & -0.69     & -0.21          & 0.21            & -0.69 \\ 
    \hline                     
 \multirow {4}*{S6} & $\uparrow \uparrow$            & 0.69       & 0.17            & 0.22           & 0.30  \\ 
                                & $\downarrow \downarrow$  & 0.20       & -0.20          & -0.17          & 0.57  \\ 
                                & $\uparrow \downarrow$       & 0.45       & 0.18            & -0.17           & 0.44  \\ 
                                & $\downarrow \uparrow$       & 0.69       & -0.22           & 0.22           & 0.69 \\ 
   \hline
   \hline
    \end{tabular}
\label{table1}
\end{table}

\subsection{Switching of polarization multistates}

\begin{figure*}
\centering
\includegraphics[width=.95\linewidth]{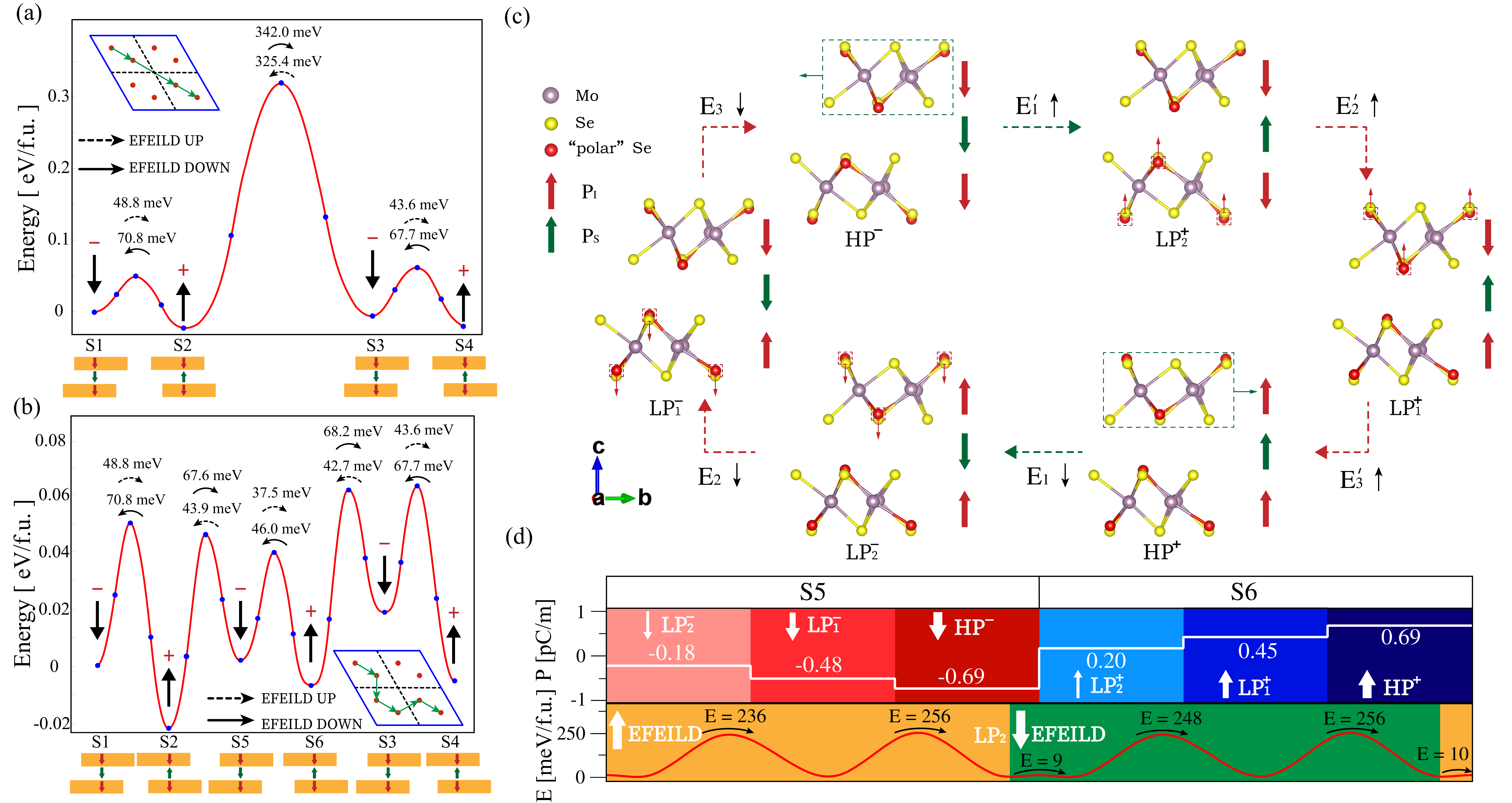}
\caption{Transformation of the polarization multistates in a 1T"-MoSe$_2$ bilayer. (a) and (b) Energy barriers for the pathways along S1-S2-S3-S4 and S1-S2-S5-S6-S3-S4, respectively (see the insets). Here we only show the barriers for the layers with downward intrinsic polarizations. Others show the same trend. (c) Switching of the sextuple polarization states for a 1T"-MoSe$_2$ bilayer. (d) Net polarizations for the multistates and energy barriers for the pathways between them.}
\label{fig3}
\end{figure*}
   
We first investigate the pathways for layer sliding. The energy barriers between different stackings are shown in Figs.~\ref{fig3}(a) and (b), where all the layers have downward polarization. The reason for keeping the intralayer polarizations unchanged is that the energy barrier for layer sliding is much smaller than that for reversing the intralayer polarizations. Assuming the system is in the state S1 with downward intralayer and interlayer polarizations, applying a small but sufficiently large electric field to reverse the stacking-induced polarization $P_S$ will drive it into S2.  There are three pathways for S2 to transform when $P_S$ is flipped, i.e., S2-S1, S2-S3, and S2-S5, respectively. The barrier for the path from S2 to S1 is about 71 meV/f.u.,  slightly higher than that for S2-S5.  The pathway S2-S3 has an energy barrier of about 325 meV/f.u., much larger than those for S2-S1 and S2-S5. Therefore, S2-S5 is favored when an external electric field is imposed on switching $P_S$ of S2. Then, there are two pathways with low energy barriers for S5 to transform, i.e., S5-S2 and S5-S6, respectively. The energy barriers for them are about 44 and 38 meV/f.u., respectively. Thus, S5 will transform into S6 rather than S2 when an electric field is applied to switch the interlayer polarization $P_S$. For S6, the energy barriers are about 46 and 68  meV/f.u. for the pathways to S5 and S3, respectively. The trends in the energy barrier for the discussed pathways suggest that the system prefers to be transformed between S5 and S6 as $P_S$ is switched by sliding. 

\begin{figure*}
\centering
\includegraphics[width=.95\linewidth]{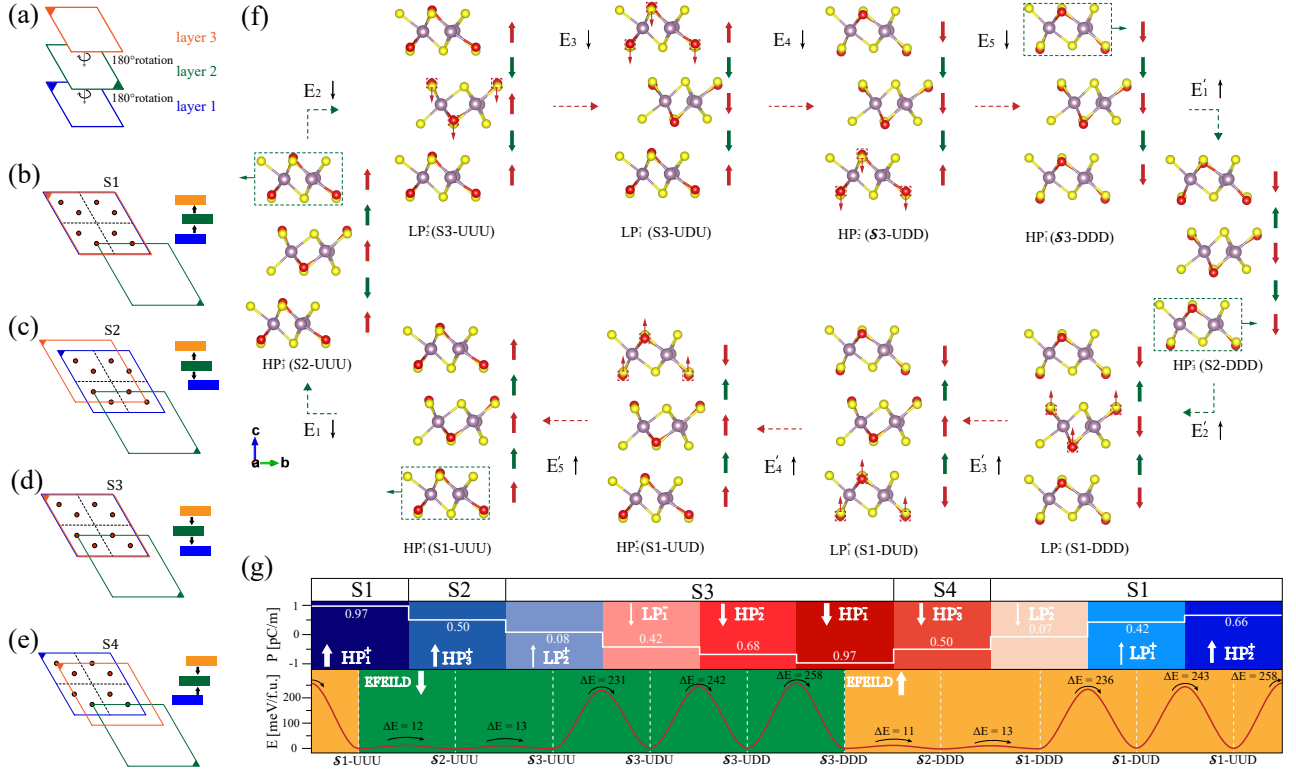}
\caption{Transformation of the polarization multistates in a 1T"-MoSe$_2$ trilayer. (a)-(e) Stacking orders involved in the transformation. (f) Schematics of switching the ten polarization states under external electric fields. (g) Magnitudes of the polarization and energy barriers for the pathways between them.}
\label{fig4}
\end{figure*}

We now discuss the transformation of the polarization multistates. We start with the state HP$^+$ [see Fig.~\ref{fig3}(c)] with upward intralayer and interlayer polarizations. An appropriate electric field opposite to the polarizations (E$_1$$\downarrow$) can flip the interlayer polarization $P_S$ by layer sliding so that the system is transformed into LP$_2^-$, for which the stacking order is S5. Then, a much larger electric field (E$_2$$\downarrow$) than E$_1$ can flip the intrinsic polarizations within the layers.  Note that the two layers are symmetrically nonequivalent. Therefore, the electric fields required to flip their polarizations are different. This difference leads to a layer-selective flipping mechanism for the intralayer polarization. Our calculations find that the energy barrier for flipping the polarization in the top layer is 10 meV/f.u. lower than that for the bottom layer of LP$_2^-$. So, the polarizations of the top layer are prior to being flipped. By this operation, the system is transformed into the state LP$_1^-$.  Then, an electric field (E$_3$$\downarrow$) larger than E$_2$$\downarrow$ can flip the polarizations of the bottom layer. The system is now in the state HP$^{-}$, which has all the intrinsic and stacking-induced polarizations parallel to the external electric field. Likewise, reversing the electric field properly can yield another three polarization states, which are named LP$_2^+$, LP$_1^+$, and HP$^+$, respectively. Therefore, there are sextuple polarization states for the bilayers of  1T"-TMD, which can be achieved by controlling the external electric field. 

Note that the energy difference in the barriers for switching the intralayer polarization $P_I$ is comparable to the barrier for the layer sliding. For instance, there is an energy difference of 10 meV/f.u. between the barriers for the pathways LP$_2^+$-LP$_1^+$ and LP$_1^+$-HP$^+$ [see Fig.~\ref{fig3}(d)]. This energy difference is the same as the barrier for the layer sliding from HP$^+$ to LP$_2^{-}$. Therefore, the difference between E$_3^{'}$ and E$_2^{'}$ is expected to be comparable to E$_1 \downarrow$. The sextuple polarizations states are also shown for the bilayer of 1T"-MoS$_2$ (see Sec. S3 in SM). It should be mentioned that there are only two (three) polarization states can be achieved for the bilayers with sliding (intrinsic) ferroelectrics. Thus, combining these two kinds of polarization can double or triple the number of states compared to those solely having one ingredient. 

One can obtain much more polarization states for thicker films than the bilayer by combining the sliding and intrinsic ferroelectricity. Fig.~\ref{fig4} shows our results for a trilayer of 1T"-MoSe$_2$. Figs.~\ref{fig4}(a)-(e) depicts the stacking orders involved in the transformation of the polarization states, in which the layers are in the H-stacking. The energy barriers for sliding the individual layers are different since they are symmetrically different. We have investigated all possible pathways for the transformation of the polarization multistates, including sliding a single layer and simultaneously sliding two layers. Our results indicate that a mechanism involving layer-by-layer sliding and switching is favorable. Assuming the system is in the HP$_1^+$ state, the energy barrier for sliding the bottom layer (12 meV/f.u.) is lower than that for sliding the top layer (13 meV/f.u.), which gives rise to a layer-selective sliding for the sliding-induced polarization. Thus, the HP$_1^+$ state transforms into HP$_3^{+}$ first and then transforms into LP$_2^{+}$ after reversing the sliding-induced polarizations. Like the bilayer, there is also a layer-selective flipping mechanism for switching the intralayer polarization of the trilayer. However, the trend is different. Instead, for the trilayer in LP$_2^{+}$, flipping the middle layer has a lower energy barrier than flipping the top or the bottom layer. The reason is that the middle layer experiences an effective external electric field due to the two interlayer polarization opposite to its intralayer polarization. Moreover, there is a difference of about 16 meV/f.u. in the energy barriers for flipping the polarizations in the top and bottom layers. As a result, a trilayer of 1T"-MoSe$_2$ possesses ten polarization states, which can be achieved by controlling the external electric field. The transformation of the polarization multistates is shown in fig.~\ref{fig4}(f), for which the energy barriers and magnitudes of the polarizations are shown in fig.~\ref{fig4}(g). According to the trend in the energy barrier, the electric fields required for reversing the polarization during the transformation have the trend $E_1 \downarrow$ $<$ $E_2 \downarrow$ $\ll$  $E_3 \downarrow$ $<$ $E_4 \downarrow$ $<$ $E_5 \downarrow$.  

\section{Conclusions}
In summary,  we have proposed to build ferroelectric multistates by combining the intrinsic and sliding-induced polarizations in two-dimensional materials. We have illustrated the concept in 1T"-TMD bilayers and trilayers using first-principles calculations. Our NEB calculations reveal that a mechanism involving layer-by-layer sliding and flipping can give rise to six and ten switchable states by controlling the external electric field. Our concept is general and can be applied to other 2D ferroelectrics with similar features. The polarization multistates revealed in this study not only can significantly increase the information storage density but also provide opportunities for designing novel electronic devices and manipulating the electronic properties of overlayers in heterostructures. 

\begin{acknowledgments}
This work was supported by the National Natural Science Foundation of China (Grants Nos. 12174098,  11774084, and No. 52372260), the Science Fund for Distinguished Young Scholars of Hunan Province of China (Grant No. 2024JJ2048), and the Youth Science and Technology Talent Project of Hunan Province (Grant No. 2022RC1197), and the State Key Laboratory of Powder Metallurgy, Central South University, Changsha, China. Calculations were carried out in part using computing resources at the High-Performance Computing Platform of Hunan Normal University.
\end {acknowledgments}

\bibliography{references}
\bibliographystyle{apsrev4-2}

\end{document}